\documentclass[prc,preprint,showpacs,preprintnumbers,amsmath,amssymb]{revtex4}

\usepackage[mathscr]{eucal}
\usepackage{graphicx}
\def\slr#1{\setbox0=\hbox{$#1$}           
   \dimen0=\wd0                                 
   \setbox1=\hbox{/} \dimen1=\wd1               
   \ifdim\dimen0>\dimen1                        
      \rlap{\hbox to \dimen0{\hfil/\hfil}}      
      #1                                        
   \else                                        
      \rlap{\hbox to \dimen1{\hfil$#1$\hfil}}   
      /                                         
   \fi}

\def\ksq{k^2}

\def\gev#1{ GeV${}^{#1}$}
\def\be{\begin{eqnarray}}
\def\ee{\end{eqnarray}}

\renewcommand{\theequation}%
    {\arabic{section}.\arabic{equation}}
\makeatletter \@addtoreset{equation}{section} \makeatother

\begin{document}

\preprint{BCCNT: 04/05/321}

\title{Calculation of the Momentum Dependence of Hadronic
Current Correlation Functions at Finite Temperature}

\author{Xiangdong Li}
\affiliation{%
Department of Computer System Technology\\
New York City College of Technology of the City University of New
York\\
Brooklyn, New York 11201 }%

\author{Hu Li}
\author{C. M. Shakin}
\email[email address:]{casbc@cunyvm.cuny.edu}
\author{Qing Sun}
\author{Huangsheng Wang}

\affiliation{%
Department of Physics and Center for Nuclear Theory\\
Brooklyn College of the City University of New York\\
Brooklyn, New York 11210
}%

\date{May, 2004}

\begin{abstract}
We have calculated spectral functions associated with hadronic
current correlation functions for vector currents at finite
temperature. We made use of a model with chiral symmetry,
temperature-dependent coupling constants and temperature-dependent
momentum cutoff parameters. Our model has two parameters which are
used to fix the magnitude and position of the large peak seen in
the spectral functions. In our earlier work, good fits were
obtained for the spectral functions that were extracted from
lattice data by means of the maximum entropy method (MEM). In the
present work we extend our calculations and provide values for the
three-momentum dependence of the vector correlation function at
$T=1.5\,T_c$. These results are used to obtain the correlation
function in coordinate space, which is usually parametrized in
terms of a screening mass. Our results for the three-momentum
dependence of the spectral functions are similar to those found in
a recent lattice QCD calculation for charmonium [S. Datta, F.
Karsch, P. Petreczky and I. Wetzorke, hep-lat/0312037]. For a
limited range we find the exponential behavior in coordinate space
that is usually obtained for the spectral function for $T>T_c$ and
which allows for the definition of a screening mass.
\end{abstract}


\maketitle

\section{INTRODUCTION}
In a number of recent works [1-\,3] we have calculated various
hadronic correlation functions and compared our results to results
obtained in lattice simulations of QCD [4-\,6]. The lattice
results for the correlators, $G(\tau, T)$, may be used to obtain
the corresponding spectral functions, $\sigma(\omega, T)$, by
making use of the relation \be G(\tau, T)=\int_0^\infty d \omega
\sigma(\omega, T) K(\tau, \omega, T)\,,\ee where \be K(\tau,
\omega, T)=\frac{\cosh[\omega(\tau-1/2T)]}{\sinh(\omega/2T)}\,.\ee
The procedure to obtain $\sigma(\omega, T)$ from the knowledge of
$G(\tau, T)$ makes use of the maximum entropy method (MEM) [7-9],
since $G(\tau, T)$ is only known at a limited number of points.

In our studies of meson spectra at $T=0$ and at $T<T_c$ we have
made use of the Nambu--Jona-Lasinio (NJL) model. The Lagrangian of
the generalized NJL model we have used in our studies is

\begin{flushleft}
\be \mathcal L&=&\overline{q}(i\slr\gamma-m^0)q+\frac{
G_S}{2}\sum_{i=0}^8 [(\overline{q} \lambda^{i} q)^2+(\overline{q}
i \gamma_5 \lambda^{i} q)^2]\\\nonumber &-&\frac{
G_V}{2}\sum_{i=0}^{8}[(\overline{q} \lambda^{i}\gamma_\mu
q)^2+(\overline{q} \lambda^{i}\gamma_5\gamma_\mu q)^2]\\\nonumber
&+&\frac{G_D}{2} \lbrace
\det[\overline{q}(1+\lambda_5)q]+\det[\overline{q}(1-\lambda_5)q]\rbrace
+\mathcal L_{conf}\,. \ee
\end{flushleft}

Here, $m^0$ is a current quark mass matrix, $m^0$=diag$(m_u^0,
m_d^0, m_s^0)$. The $\lambda_i$ are the Gell-Mann (flavor)
matrices and $\lambda^0=\sqrt{2/3}\mathbf{1}$, with $\mathbf{1}$
being the unit matrix. The fourth term is the 't Hooft interaction
and $\mathcal L_{conf}$ represents the model of confinement used
in our studies of meson properties.

In the study of hadronic current correlators it is important to
use a model which respects chiral symmetry, when $m^0=0$.
Therefore, we make use of the Lagrangian of Eq. (1.3), while
neglecting the 't Hooft interaction and $\mathcal L_{conf}$. Thus,
there are essentially three parameters to consider, $G_S$, $G_V$
and a cutoff parameter $k_{max}$, which restricts the momentum
integrals so that $k<k_{max}$. When we use the NJL model to study
matter at finite temperature, we introduce the
temperature-dependent parameters $G_S(T)$ and $G_V(T)$. (We have
also used a Gaussian cutoff for the momentum integrals in our
earlier work.) These parameters have been adjusted to obtain fits
to the spectral functions $\sigma(\omega, T)$ for $T/T_c=1.5$ and
$T/T_c=3.0$, which are the values of $T/T_c$ studied in the
lattice simulations of QCD [10].

The temperature-dependent coupling constants and cutoff parameters
of our work are analogous to the corresponding density-dependent
parameters introduced in Ref. [11] and [12]. Further study of
models with temperature-dependent and density-dependent parameters
are of interest and a general theoretical formalism for the
introduction of such dependencies should be considered.

In this work we limit our study to the data for the vector
correlator at $T=1.5\,T_c$ [10, 13] and therefore only need to
specify $G_V(T)$ and the momentum cutoff. (We remark that the
results for the scalar, vector, pseudoscalar and axialvector
correlators are quite similar in the deconfined region.) In Figs.
1 and 2 we show the data obtained by the MEM method at $T/T_c=1.5$
and $T/T_c=3.0$ for both pseudoscalar and vector correlators [10,
13]. The second peaks in these correlators are known to be a
lattice artifacts [13].

The organization of our work is as follows. In Appendix A we
present the formalism for calculation of pseudoscalar and vector
correlation functions. In Appendix A we discuss the calculation of
the correlator in the case the quark and antiquark carry zero
total momentum. In Appendix B we show how the formalism is
modified for the correlator calculated at finite momentum
$\overrightarrow P$. In Section II we present the results of our
calculation of the imaginary part of the correlator
$\sigma(\omega, \overrightarrow P)$. (Since we place
$\overrightarrow P$ along the \emph{z}-axis this quantity may be
written as $\sigma(\omega, 0, 0, P_z)$ in accord with the notation
of Ref. [10].) Our results for $\sigma(\omega, \overrightarrow P)$
will be presented for a series of values of $|\overrightarrow P|$
in Section II. In Section III we present our result for the
coordinate-dependent correlator $C(z)$ which is proportional to
the correlator defined in Eq. (1) of Ref. [10], \be
C(z)=\frac12\int_{-\infty}^\infty\,dP_ze^{iP_zz}\int_0^\infty\,d\omega\frac{\sigma(\omega,
0, 0, P_z)}\omega\,. \ee We may also use the form \be
C(z)=\frac14\int_{-\infty}^\infty\,dP_ze^{iP_zz}\int_0^\infty\,dP^2\,\frac{\sigma(P^2,
0, 0, P_z)}{P^2}\,. \ee Finally, in Section IV we present our
conclusion and further discussion.

\section{temperature dependent hadronic current correlators at finite
momentum}

We make use of the formalism presented in Appendices A and B to
obtain values of the vector correlator at $T=1.5\,T_c$. Here we
take $T_c=270$ MeV, since we have usually made comparison to
lattice calculations made in the quenched approximation.

In Fig. 3 we present $\sigma(\omega)/\omega^2$, for various values
of $|\overrightarrow P|$, as function of $\omega^{2}$. Comparison
may be made to the lattice data shown in Fig. 2 [10, 13]. (We
again note that our calculation does not reproduce the second peak
in the lattice data which is known to be a lattice artifact [13].)
The curves shown in Fig. 3 are given for values of
$|\overrightarrow P|$ ranging from 0.10 GeV to 2.10 GeV in steps
of 0.20 GeV. For these calculations we have used
$k_{max}=1.22$\,GeV. Our results for the various values of
$|\overrightarrow P|$ given in Fig.\,3 may be compared to Fig.\,20
of Ref.\,[14]. We see that the results calculated by completely
different methods are similar.

\begin{figure}
\includegraphics[bb=0 0 280 235, angle=0, scale=1]{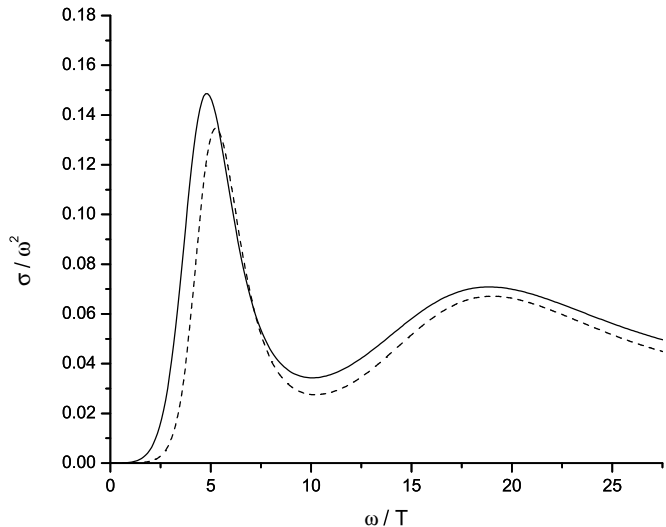}%
\caption{The spectral functions $\sigma/\omega^2$ for pseudoscalar
states obtained by MEM are shown [10,\,13]. The solid line is for
$T/T_c=1.5$ and the dashed line is for $T/T_c=3.0$. The second
peak is a lattice artifact [13].}
\end{figure}

\begin{figure}
\includegraphics[bb=0 0 280 235, angle=0, scale=1]{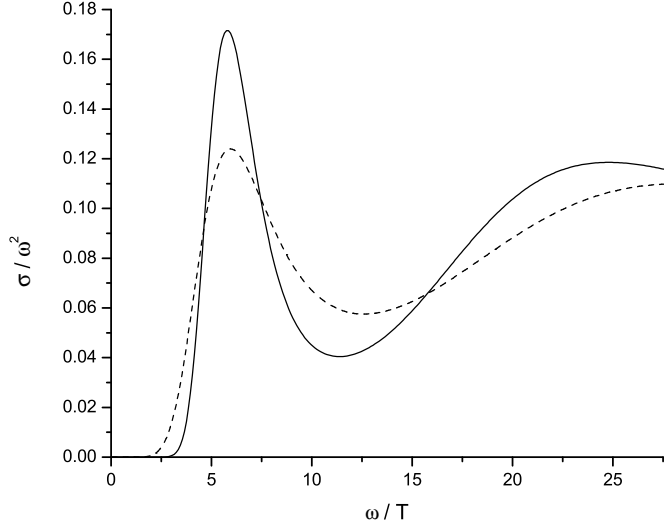}%
\caption{The spectral functions $\sigma/\omega^2$ for vector
states obtained by MEM are shown [10, 13]. See the caption of Fig.
1. The second peak is a lattice artifact [13].}
\end{figure}

\begin{figure}
\includegraphics[bb=0 0 280 235, angle=0, scale=1]{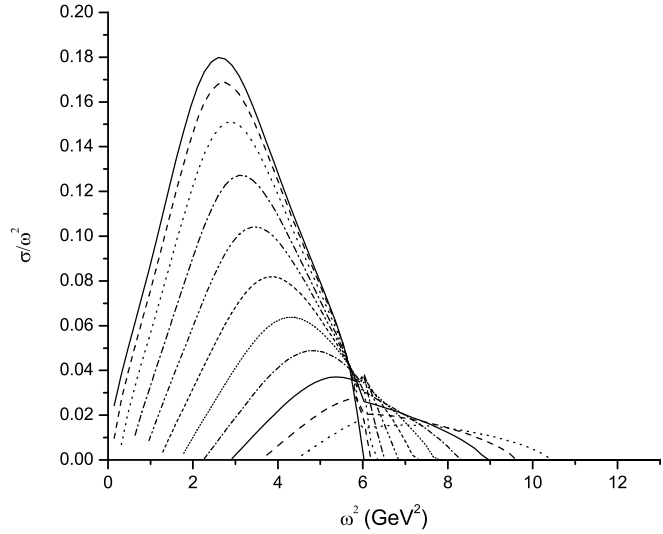}%
\caption{The imaginary part of the correlator
$\sigma(\omega)/\omega^2$ is shown for various values of $|\vec
P|$ as a function of $\omega^{2}$. Starting with the topmost curve
the values of $|\vec P|$ in GeV units are 0.10, 0.30, 0.50, 0.70,
0.90, 1.10, 1.30, 1.50, 1.70, 1.90 and 2.10. Here we have used
$G_S=1.2$ GeV$^{-2}$ and $k_{max}=1.22$ GeV.}
\end{figure}

\begin{figure}
\includegraphics[bb=0 0 280 235, angle=0, scale=1]{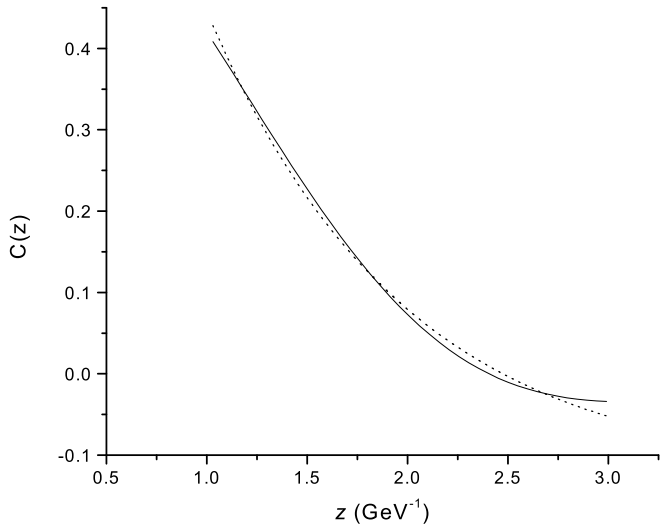}%
\caption{The correlation function $C(z)$ defined in Eq.\,(1.5) is
shown. The dotted line represents a fit using an exponential
function.}
\end{figure}

\section{the coordinate-space correlation function}

We have used the results of our calculations which were presented
in Fig. 3 to calculate $C(z)$ of Eq. (1.5). The result of that
calculation is shown in Fig. 4. We note that the simple assumption
for the behavior of this correlator that is usually made,
$C(z)\thicksim\exp[-m_{sc}z]$, is born out in our calculation for
1\,GeV$^{-1}<\,z<\,3\,$GeV$^{-1}$. For the study of charmonium on
the lattice, values found for the screening mass are given in Ref.
[14].

For the result shown in Fig. 4 we obtain a screening mass of 1.02
GeV which may be compared to the value of $\pi$$T$=1.27 GeV. The
dotted line represents an exponential fit to our result.

\section{discussion}

Recent theoretical work concerning the quark-gluon plasma has been
discussed by Shuryak \cite{Sh03a}. He notes that the physics of
excited matter produced in heavy-ion collisions in the region
$T_c<T<4T_c$ is different from that of a weakly coupled
quark-gluon plasma. That is due to the strong coupling generated
by the bound states of the quasiparticles. These bound states
appear as resonant structures when the imaginary parts of the
hadronic correlators are extracted from lattice data using the
maximum entropy method (MEM) [4-9]. These resonances lead to very
strong interactions between the quasiparticles and are, in part,
responsible for very small mean free paths and collective flow in
heavy-ion collisions. That flow may be described by hydrodynamics.
Indeed, Shuryak suggests that ``... if the system is
macroscopically large, then its description via
\emph{thermodynamics} of its bulk properties (like matter
composition) and \emph{hydrodynamics} for space-time evolution
should work" \cite{Sh03a}.

In our present work we have presented a simple chiral model that
is able to reproduce the resonances extracted from the lattice
data via the MEM procedure. We have calculated the imaginary parts
of the vector correlator for various values of the total external
momentum $\overrightarrow P$. We have also calculated the
correlation function, $C(z)$, and find the simple exponential
behavior $C(z)\thicksim\exp[-m_{sc}z]$ for a limited range of
\emph{z}. We have shown in an earlier work that exponential
behavior with the appropriate screening mass may be obtained for
the full range of \emph{z} values if quite small values of
$k_{max}$ (of the order of 0.4 GeV) are used \cite{Li04a}.
Finally, we note that an extensive discussion of screening masses
appears in Ref. \cite{Fl97a}.

\appendix
  \renewcommand{\theequation}{A\arabic{equation}}
  \setcounter{equation}{0}  
  \section{}  

For ease of reference, we present a discussion of our calculation
of hadronic current correlators taken from Ref.\,[3]. The
procedure we adopt is based upon the real-time finite-temperature
formalism, in which the imaginary part of the polarization
function may be calculated. Then, the real part of the function is
obtained using a dispersion relation. The result we need for this
work has been already given in the work of Kobes and Semenoff
\cite{Ko85a}. (In Ref. \cite{Ko85a} the quark momentum is $k$ and
the antiquark momentum is $k-P$. We will adopt that notation in
this section for ease of reference to the results presented in
Ref. \cite{Ko85a}.) With reference to Eq.\,(5.4) of Ref.
\cite{Ko85a}, we write the imaginary part of the scalar
polarization function as \be \mbox{Im}\,J_S(\textit{P}\,{}^2,
T)=\frac12N_c\beta_S\,\epsilon(\textit{P}\,{}^0)\int^{k_{max}}
\frac{d^{3}k}{(2\pi)^{3}}\left(\frac{2\pi}{2E_1(k)2E_2(k)}\right)\\\nonumber
\times\{[1-n_1(k)-n_2(k)]
\delta(\textit{P}\,{}^0-E_1(k)-E_2(k))\\\nonumber-[n_1(k)-n_2(k)]
\delta(\textit{P}\,{}^0+E_1(k)-E_2(k))\\\nonumber-[n_2(k)-n_1(k)]
\delta(\textit{P}\,{}^0-E_1(k)+E_2(k))\\\nonumber-[1-n_1(k)-n_2(k)]
\delta(\textit{P}\,{}^0+E_1(k)+E_2(k))\}\,.\ee Here, $E_1(k)=[\vec
k\,{}^2+m_1^2(T)]^{1/2}$. Relative to Eq.\,(5.4) of Ref.
\cite{Ko85a}, we have changed the sign, removed a factor of $g^2$
and have included a statistical factor of $N_c$. In addition, we
have used a sharp cutoff, $k_{max}$, for the momentum integral. We
also note that \be n_1(k)=\frac1{e^{\,\beta E_1(k)}+1}\,,\ee and
\be n_2(k)=\frac1{e^{\,\beta E_2(k)}+1}\,.\ee For the calculation
of the imaginary part of the polarization function, we may put
$\ksq=m_1^2(T)$ and $(k-P)^2=m_2^2(T)$, since in that calculation
the quark and antiquark are on-mass-shell. In Eq.\,(A1) the factor
$\beta_S$ arises from a trace involving Dirac matrices, such that
\be \beta_S&=&-\mbox{Tr}[(\slr k+m_1)(\slr k-\slr P+m_2)]\\
&=&2P^2-2(m_1+m_2)^2\,,\ee where $m_1$ and $m_2$ depend upon
temperature. In the frame where $\vec P=0$, and in the case
$m_1=m_2$, we have $\beta_S=2P_0^2(1-{4m^2}/{P_0^2})$. For the
scalar case, with $m_1=m_2$, we find \be \mbox{Im}\,J_S(P^2,
T)=\frac{N_cP_0^2}{8\pi}\left(1-\frac{4m^2(T)}{P_0^2}\right)^{3/2}
[1-2n_1(k)]\,,\ee where \be \vec
k\,{}^2=\frac{P_0^2}4-m^2(T)\,,\ee with $k<k_{max}$.

For pseudoscalar mesons, we replace $\beta_S$ by \be
\beta_P&=&-\mbox{Tr}[i\gamma_5(\slr k+m_1)i\gamma_5(\slr k-\slr
P+m_2)]\\
&=&2P^2-2(m_1-m_2)^2\,,\ee which for $m_1=m_2$ is $\beta_P=2P_0^2$
in the frame where $\vec P=0$. We find, for the $\pi$ mesons, \be
\mbox{Im}\,J_P(P^2,T)=\frac{N_cP_0^2}{8\pi}\left(1-\frac{4m^2(T)}{P_0^2}\right)^{1/2}
[1-2n_1(k)]\,,\ee where $ \vec k\,{}^2={P_0^2}/4-m_u^2(T)$, as
above, with $k<k_{max}$. Thus, we see that the phase space factor
has an exponent of 1/2 corresponding to a \textit{s}-wave
amplitude. For the scalars, the exponent of the phase-space factor
is 3/2, as seen in Eq.\,(A6).

For a study of vector mesons we consider \be
\beta_{\mu\nu}^V=\mbox{Tr}[\gamma_\mu(\slr k+m_1)\gamma_\nu(\slr
k-\slr P+m_2)]\,,\ee and calculate \be
g^{\mu\nu}\beta_{\mu\nu}^V=4[P^2-m_1^2-m_2^2+4m_1m_2]\,,\ee which,
in the equal-mass case, is equal to $4P_0^2+8m^2(T)$, when
$m_1=m_2$ and $\vec P=0$. This result will be needed when we
calculate the correlator of vector currents. Note that, for the
elevated temperatures considered in this work, $m_u(T)=m_d(T)$ is
quite small, so that $4P_0^2+8m_u^2(T)$ can be approximated by
$4P_0^2$, when we consider the vector current correlation
functions. In that case, we have \be \mbox{Im}\,J_V(P^2,T) \simeq
\frac{2}{3}\mbox{Im}\,J_P(P^2,T)\,.\ee At this point it is useful
to define functions that are not for $k>k_{max}$:
\be\mbox{Im}\,\tilde{J}_P(P^2,T)=\frac{N_cP_0^2}{8\pi}\left(1-\frac{4m^2(T)}{P_0^2}\right)^{1/2}[1-2n_1(k)]\,,\ee
and
\be\mbox{Im}\,\tilde{J}_V(P^2,T)=\frac{2}{3}\frac{N_cP_0^2}{8\pi}\left(1-\frac{4m^2(T)}{P_0^2}\right)^{1/2}[1-2n_1(k)]\,,\ee
For the functions defined in Eq.\,(A14) and (A15) we need to use a
twice-subtracted dispersion relation to obtain
$\mbox{Re}\,\tilde{J}_P(P^2,T)$, or
$\mbox{Re}\,\tilde{J}_V(P^2,T)$. For example,
\be\mbox{Re}\,\tilde{J}_P(P^2,T)=\mbox{Re}\,\tilde{J}_P(0,T)+
\frac{P^2}{P_0^2}[\mbox{Re}\,\tilde{J}_P(P_0^2,T)-\mbox{Re}\,\tilde{J}_P(0,T)]\\\nonumber
+\frac{P^2(P^2-P_0^2)}{\pi}\int_{4m^2(T)}^{\tilde{\Lambda}^{2}}
ds\frac{\mbox{Im}\,\tilde{J}_P(s,T)}{s(P^2-s)(P_0^2-s)}\,,\ee
where $\tilde{\Lambda}^{2}$ can be quite large, since the integral
over the imaginary part of the polarization function is now
convergent. We may introduce $\tilde{J}_P(P^2,T)$ and
$\tilde{J}_V(P^2,T)$ as complex functions, since we now have both
the real and imaginary parts of these functions. We note that the
construction of either $\mbox{Re}\,J_P(P^2,T)$, or
$\mbox{Re}\,J_V(P^2,T)$, by means of a dispersion relation does
not require a subtraction. We use these functions to define the
complex functions $J_P(P^2,T)$ and $J_V(P^2,T)$.

In order to make use of Eq.\,(A16), we need to specify
$\tilde{J}_P(0)$ and $\tilde{J}_P(P_0^2)$. We found it useful to
take $P_0^2=-1.0$ \gev2 and to put $\tilde{J}_P(0)=J_P(0)$ and
$\tilde{J}_P(P_0^2)=J_P(P_0^2)$. The quantities $\tilde{J}_V(0)$
and $\tilde{J}_V(P_0^2)$ are determined in an analogous function.
This procedure in which we fix the behavior of a function such as
$\mbox{Re}\tilde{J}_V(P^2)$ or $\mbox{Re}\tilde{J}_V(P^2)$ is
quite analogous to the procedure used in Ref. \cite{Sh95a}. In
that work we made use of dispersion relations to construct a
continuous vector-isovector current correlation function which had
the correct perturbative behavior for large
$P^2\rightarrow-\infty$ and also described the low-energy
resonance present in the correlator due to the excitation of the
$\rho$ meson. In Ref. \cite{Sh95a} the NJL model was shown to
provide a quite satisfactory description of the low-energy
resonant behavior of the vector-isovector correlation function.

We now consider the calculation of temperature-dependent hadronic
current correlation functions. The general form of the correlator
is a transform of a time-ordered product of currents, \be iC(P^2,
T)=\int d^4xe^{iP\cdot x}<\!\!<T(j(x)j(0))>\!\!>\,,\ee where the
double bracket is a reminder that we are considering the finite
temperature case.

For the study of pseudoscalar states, we may consider currents of
the form $j_{P,i}(x)=\tilde{q}(x)i\gamma_5\lambda^iq(x)$, where,
in the case of the $\pi$ mesons, $i=1,2$ and $3$. For the study of
scalar-isoscalar mesons, we introduce
$j_{S,i}(x)=\tilde{q}(x)\lambda^i q(x)$, where $i=0$ for the
flavor-singlet current and $i=8$ for the flavor-octet current
\cite{Li02a}.

In the case of the pseudoscalar-isovector mesons, the correlator
may be expressed in terms of the basic vacuum polarization
function of the NJL model, $J_P(P^2, T)$. Thus, \be C_P(P^2,
T)=J_P(P^2, T)\frac{1}{1-G_{P}(T)J_P(P^2, T)}\,,\ee where $G_P(T)$
is the coupling constant appropriate for our study of $\pi$
mesons. We have found $G_P(T)=13.49$\gev{-2} by fitting the pion
mass in a calculation made at $T=0$, with $m_u = m_d =0.364$ GeV.
The result given in Eq.\,(A18) is only expected to be useful for
small $P^2$, since the Gaussian regulator strongly modifies the
large $P^2$ behavior. Therefore, we suggest that the following
form is useful, if we are to consider the larger values of $P^2$.
\be \frac{C_{P}(P^2, T)}{P^2}=\left[\frac{\tilde{J}_P(P^2,
T)}{P^2}\right] \frac{1}{1-G_P(T)J_P(P^2, T)}\,.\ee (As usual, we
put $\vec{P}=0$.) This form has two important features. At large
$P_0^2$, ${\mbox{Im}\,C_{P}(P_0, T)}/{P_0^2}$ is a constant, since
${\mbox{Im}\,\tilde{J}_{P}(P_0^2, T)}$ is proportional to $P_0^2$.
Further, the denominator of Eq.\,(A19) goes to 1 for large
$P_0^2$. On the other hand, at small $P_0^2$, the denominator is
capable of describing resonant enhancement of the correlation
function. As we have seen, the results obtained when Eq.\,(A19) is
used appear quite satisfactory. (\,We may again refer to Ref.
\cite{Sh95a}, in which a similar approximation is described.)

For a study of the vector-isovector correlators, we introduce
conserved vector currents $j_{\mu,
i}(x)=\tilde{q}(x)\gamma_{\mu}\lambda_i q(x)$ with i=1, 2 and 3.
In this case we define \be J_V^{\mu\nu}(P^2,
T)=\left(g\,{}^{\mu\nu}-\frac{P\,{}^\mu
P\,{}^\nu}{P^2}\right)J_V(P^2, T)\ee and \be C_V^{\mu\nu}(P^2,
T)=\left(g\,{}^{\mu\nu}-\frac{P\,{}^\mu
P\,{}^\nu}{P^2}\right)C_V(P^2, T)\,,\ee taking into account the
fact that the current $j_{\mu,\,i}(x)$ is conserved. We may then
use the fact that \be J_V(P^2,T) =
\frac13g_{\mu\nu}J_V^{\mu\nu}(P^2,T)\ee and
\be\mbox{Im}\,J_V(P^2,T)&=&
\frac23\left[\frac{P_0^2+2m_u^2(T)}{8\pi}\right]
\left(1-\frac{4m_u^2(T)}{P_0^2}\right)^{1/2}[1-2n_1(k)]\\
&\simeq& \frac{2}{3}\mbox{Im}J_P(P^2,T)\,.\ee (See Eq.\,(A7) for
the specification of $k=|\vec k|$.) We then have \be
C_V(P^2,T)=\tilde{J}_V(P^2,T)\frac1{1-G_V(T)J_V(P^2,T)}\,,\ee
where we have introduced \be\mbox{Im}\tilde{J}_V(P^2,T)&=&
\frac23\left[\frac{P_0^2+2m_u^2(T)}{8\pi}\right]
\left(1-\frac{4m_u^2(T)}{P_0^2}\right)^{1/2}[1-2n_1(k)]\\
&\simeq& \frac{2}{3}\mbox{Im}\tilde{J}_P(P^2,T)\,. \ee In the
literature, $\omega$ is used instead of $P_0$ [4-6]. We may define
the spectral functions \be\sigma_V(\omega,
T)=\frac{1}{\pi}\,\mbox{Im}\,C_V(\omega, T)\,,\ee and
\be\sigma_P(\omega, T)=\frac{1}{\pi}\,\mbox{Im}\,C_P(\omega,
T)\,,\ee

Since different conventions are used in the literature [4-6], we
may use the notation $\overline{\sigma}_P(\omega, T)$ and
$\overline{\sigma}_V(\omega, T)$ for the spectral functions given
there. We have the following relations: \be
\overline{\sigma}_P(\omega, T)=\sigma_P(\omega, T)\,,\ee and
\be\frac{\overline{\sigma}_V(\omega,
T)}{2}=\frac{3}{4}\sigma_V(\omega, T)\,,\ee where the factor 3/4
arises because, in Refs. [4-6], there is a division by 4, while we
have divided by 3, as in Eq.\,(A22).

\section{}
\renewcommand{\theequation}{B\arabic{equation}}

Here we extend the work of Appendix A to consider case of finite
three-momentum, $\vec{P}$. We consider the calculation of
$\mbox{Im}J_P(P^0,\vec{P},T)$. The momenta $P^0$ and $\vec{P}$ are
the values external to the loop diagram. Internal to the diagram,
we have a quark of momentum $k+P/2$ leaving the left-hand vertex
and an antiquark of momentum $k-P/2$ entering the left-hand
vertex. It is useful to define \be
E_1(k)&=&\left|\vec{k}+\vec{P}/2\right|\\
&=&\left(k^2+\frac{P^2}4+kP\cos\theta\right)^{1/2}\ee and
\be E_2(k)&=&\left|\vec{k}-\vec{P}/2\right|\\
&=&\left(k^2+\frac{P^2}4-kP\cos\theta\right)^{1/2}\,.\ee Here
$k=|\vec{k}|$ and $P=|\vec{P}|$.

We have \be \mbox{Im}\,J_V(P^0,\vec{P},
T)=\frac12N_c\beta_V\,\epsilon(P^0)\int^{k_{max}}
\frac{d^{3}k}{(2\pi)^{3}}\left(\frac{2\pi}{2E_1(k)2E_2(k)}\right)\\\nonumber
\times\{[1-n_1(k)-n_2(k)]
\delta(P\,{}^0-E_1(k)-E_2(k))\\\nonumber-[n_1(k)-n_2(k)]
\delta(P\,{}^0+E_1(k)-E_2(k))\\\nonumber-[n_2(k)-n_1(k)]
\delta(P\,{}^0-E_1(k)+E_2(k))\\\nonumber-[1-n_1(k)-n_2(k)]
\delta(P\,{}^0+E_1(k)+E_2(k))\}\,.\ee Here, \be
n_1(k)=\frac1{e^{\,\beta E_1(k)}+1}\,,\ee and \be
n_2(k)=\frac1{e^{\,\beta E_2(k)}+1}\,.\ee In Eq. (B5), the second
and third terms cancel and the fourth term does not contribute. It
is useful to rewrite $\delta(P^0-E_1(k)-E_2(k))$ using \be
\delta[f(\cos\theta)]=\frac2{\left|\frac{\partial f}{\partial \cos
\theta}\right|_x}\delta(\cos\theta-x)\,,\ee where \be
x^2&=&\cos^2\theta\\&=&\frac{4P_0^2(k^2+P^2/4)-P_0^4}{4k^2P^2}\nonumber\,.\ee
We find \be \left|\frac{\partial f}{\partial \cos
\theta}\right|=\frac12kP\left|\frac{E_1(k)-E_2(k)}{E_1(k)E_2(k)}\right|\,,\ee
and obtain \be \mbox{Im}\,J_P(P^0,\vec{P},
T)=\frac12N_c\beta_P\,\epsilon(P^0)(2\pi)^2\int^{k_{max}}
\frac{k^2dk}{(2\pi)^3}\\\nonumber
\int\frac1{2E_1(k)E_2(k)}[1-n_1(k)-n_2(k)]
\left|\frac{\partial{f(\cos\theta)}}{\partial{
\cos\theta}}\right|\\\nonumber
\times\delta(\cos\theta-x)d(\cos\theta)\,.\ee We note there is a
singularity when $E_1(k)=E_2(k)$. That occurs when $\cos\theta=0$
or $\theta=\pi/2$. For our calculations we eliminate the point
with $\theta=\pi/2$ when evaluating the angular integral over
$d(\cos\theta)\delta(\cos\theta-x)$ in the last expression. We
obtain \be \mbox{Im}\,J_P(P^0,\vec{P},
T)=N_c\beta_P\,\epsilon(P^0)\frac{4\pi^2}{(2\pi)^3}\int^{k_{max}}
k^2dk\,\\\nonumber
\times\left.\frac{1-n_1(k)-n_2(k)}{kP|E_1(k)-E_2(k)|}\right|_x\,,\ee
where \emph{x} is obtained from Eq. (B9), \be
x=\frac{P^0}{kP}\left[k^2+\frac{P^2}4-\frac{P_0^2}4\right]^{1/2}\ee

\begin{acknowledgments}
We wish to thank Peter Petreczky for suggesting the work reported
here and for his explaining various features of the relevant
lattice calculations.
\end{acknowledgments}



\vspace{1.5cm}
\begin{thebibliography}{99}
    \bibitem{B1}Bing He, Hu Li, C. M. Shakin, and Qing Sun, Phys. Rev. D
\textbf{67}, 014022 (2003).
    \bibitem{B2}Bing He, Hu Li, C. M. Shakin, and Qing Sun, Phys. Rev. D
\textbf{67}, 114012 (2003).
    \bibitem{B3}Bing He, Hu Li, C. M. Shakin, and Qing Sun, Phys. Rev. C
\textbf{67}, 065203 (2003).
    \bibitem{B4}I. Wetzorke, F. Karsch, E. Laermann, P. Petreczky, and S.
Stickan, Nucl. Phys. B (Proc. Suppl.) \textbf{106}, 510 (2002)
    \bibitem{B5}F. Karsch, S. Datta, E. Laermann, P. Petreczky, and S.
Stickan, and I. Wetzorke, Nucl. Phys. A \textbf{715}, 701c (2003)
    \bibitem{B6}F. Karsch, E.Laermann, P. Petreczky, S. Stickan, and I.
Wetzorke, Phys. Lett. B \textbf{530}, 147 (2002).
    \bibitem{B7}M. Asakawa, T. Hatsuda and Y. Nakahara, Nucl. Phys. A
\textbf{715}, 863 (2003)
    \bibitem{B8}T. Umeda, K. Nomura and H. Matsufuru, hep-ph/0211003.
    \bibitem{B9}I. Wetzorke, hep-ph/0305012.
(Invited talk at the 'Seventh Workshop on Quantum Chromodunamics',
Villefranche-sur-mer, France, Jan. 6-10, 2003)
    \bibitem{B10}P. Petreczky, J. Phys.\;G \textbf{30}, S431-S440 (2004).
    \bibitem{B11}M. Ruggieri, hep-ph/0310145.
    \bibitem{B12}R. Casalbuoni, R. Gatto, G. Nardulli, and M. Ruggieri,
Phys. Rev. D \textbf{68}, 034024 (2003).
    \bibitem{B13}P. Petreczky, private communication.
    \bibitem{B14}S. Datta, F. Karsch, P. Petreczky and I. Wetzorke, hep-lat/0312037.
    \bibitem{Da97a}A. Das, \emph{Finite Temperature Field Theory}
    (World Scientific, Singapore, 1997).
    \bibitem{Sh03a}E. Shuryak, hep-ph/0312227.
    \bibitem{Ko85a}R. L. Kobes and G. W. Semenoff, Nucl. Phys. B \textbf{260},
714 (1985).
    \bibitem{Sh95a}C. M. Shakin, Wei-Dong Sun, and J. Szweda, Ann. of Phys.
(NY) \textbf{241}, 37 (1995).
    \bibitem{Li02a}Hu Li and C.M. Shakin, hep-ph/0209136.
    \bibitem{Li04a}Xiangdong Li, Hu Li, C.M. Shakin, and Qing Sun,
    nucl-th/0405035.
    \bibitem{Fl97a}W. Florkowski, Acta Phys. Polon. B \textbf{28},
    2079 (1997).


\end{thebibliography}

\end{document}